\documentclass[prb,rsi,amsmath,amssymb,reprint]{revtex4-1}

\usepackage{graphicx}
\usepackage{dcolumn}
\usepackage{bm}
\usepackage{natbib}
\setcitestyle{square}

\usepackage{amsmath}
\usepackage{braket}
\usepackage{mathrsfs}
\usepackage{amsfonts}
\usepackage{color}
\usepackage[normalem]{ulem} 
\usepackage{cellspace}%
\usepackage{hyperref}
\usepackage{siunitx}
\usepackage{float}

\hypersetup{
    colorlinks=true,
    linkcolor=blue,
    filecolor=magenta,      
    urlcolor=black,
    citecolor=red,
}

\newcommand{\angstrom}{\mbox{\normalfont\AA}}

\begin{document}


\title{Electron-electron Scattering in the Diffusive Regime and Dyakonov-Perel Theory: a case study of spin decoherence in {\it n}-type  GaAs} 


\author{Gionni Marchetti}
\email{gionnimarchetti@ub.edu}

\affiliation{%
Departament de F\'{i}sica de la Mat\`{e}ria Condensada, Facultat de Física,  Universitat de Barcelona, Carrer Mart\'{i} i Franqu\`{e}s 1, 08028, Barcelona, Spain\\
Institut de Nanoci{\`e}ncia i Nanotecnologia, Universitat de Barcelona, Av. Joan XXIII S/N, 08028, Barcelona, Spain}




\date{\today}

\begin{abstract}

We argue that in the ensemble Monte Carlo approach, the spin decoherence caused by electron-electron interactions in {\it n}-type bulk GaAs at room temperature, according to the Dyakonov-Perel theory, is mainly induced by the electron density while electron-electron scattering itself  does not directly affect spin lifetime but influence it  through any energy dependence of other scattering mechanisms as typically expected in the case of the electron mobility. Therefore, our analysis, which is  strongly supported by Bayesian Occam’s razor, reconciles the  semiclassical Monte Carlo approach with the novel computational first-principles tool recently proposed by Xu, Habib, Sundararaman, and Ping [PHYSICAL REVIEW B \textbf{104}, 184418 (2021)].
\end{abstract}

\maketitle

\section{Introduction} \label{intro}

Recently, Xu, Habib, Sundararaman, and Ping (XHSP) put forward a new  computational first-principles tool that enable the researchers to simulate ultra-fast spin dynamics in various solid-state systems~\cite{xu2021}. XHSP then went on to compute the spin relaxation time $\tau_s$ due to  Dyakonov-Perel (DP)  theory~\cite{zutic2004} in  {\it n}-type bulk  Gallium Arsenide (GaAs), a candidate material for harnessing the electron spin  in spintronic devices~\cite{zutic2004, belykh2018}, at different temperatures and doping concentrations.  In such a material, they found that the electron-electron (e-e) scattering direct contribution to $\tau_s$ is negligible. However,  by comparison of their findings with those obtained by the ensemble Monte Carlo (EMC) method ~\cite{marchetti2014, marchetti2014a} for the same  {\it n}-type  semiconductor, with  dopant concentration $n_i$ ranging from \SI{}{10^{16}\cm^{-3}} to $2.5 \times  10^{17}$ cm$^{-3}$ at $T=\SI{300}{\kelvin}$,  they argued that in the latter approach the direct contribution of e-e scattering probability to  $\tau_s$   would be greatly overestimated due to the semi-classical nature of this method~\cite{jacoboni1983}. 

Despite the fact that the EMC has some limitations, for instance it rests on  a simplified band structure (single parabolic band model)  and the phonon dispersions stem from model hamiltonians,  it still works well for simulating the e-e collisions in a diffusive regime in which the electrons move as independent particles~\cite{polini2020}. The above physical picture is indeed confirmed by the good agreement between the photoluminescence spectroscopy experiments of Oertel et al.~\cite{oertel2008} and the EMC simulations reported in Refs.~\cite{marchetti2014, marchetti2014a} (see also the inset of Fig.~\ref{fig:cross_section}). Note that in both approaches, it is reasonably assumed that the material is not a strongly correlated system. It is also found that polar optical phonon  scattering, also called Fr\"{o}hlich scattering,  dominates  spin relaxation at room temperature~\cite{marchetti2014, xu2021}, similar to the case of electron mobility.~\cite{faghaninia2015, chatto1982, fischetti2016}. Therefore, it would be truly surprising to observe a dramatic discrepancy in the direct contribution of electron-electron scattering to the spin lifetime from these two computational methods.

As a matter of course, due to the momentum-dependence of the Dyakonov-Perel mechanism, one would expect that electron-electron  scattering 
contributes directly to the spin lifetime~\cite{Glazov2003, Glazov2004}.
However, within the EMC approach, the frequency $w_{\mathrm{ee}}$ at which e-e collisions occur is accounted for by Fermi's golden rule (FGR) (see Eq.~\ref{eq:fgr}). As a result, the role of the e-e interaction in spin decoherence depends on both the electron density, denoted as $n_e$, and the actual scattering probability, measured here by the total cross-section $\sigma_{\rm ee}^{\rm B}$ in the first Born approximation (1BA)~\footnote{According to the FGR, their effects on $\tau_s$ are intertwined and, therefore, challenging to separate.}. Then, our working hypothesis is that only the electron density directly affects the spin lifetime. If this were the case, then the semiclassical and ab initio approaches would yield the same physics at room temperature.

In this Letter, we aim to tackle the aforementioned difficult problem by examining the functional forms of both the spin relaxation time and the scattering probability. To accomplish this, we will fit simple and physically motivated curves with a small number of free parameters to the respective data. In essence, our approach involves selecting the best-fit models that avoid excessive detail and over-parametrization,  in accordance with the Bayesian Occam's razor~\cite{jefferys1992, macKay1992, Ghahramani2015}. As a result, we shall demonstrate that due to e-e interactions  the spin relaxation  time grows linearly with the electron density, i.e $ \tau_{\rm ee}^{\rm EMC} \sim n_e$,   while the scattering probability decays exponentially as $ \sigma_{\rm ee} ^{\rm B} \sim  \exp(-\alpha n_e) $ where $\alpha \approx 0.06$ (see also Figs.~\ref{fig:fitTau}, ~\ref{fig:fitSigma}, respectively). When comparing these two trends, it becomes evident that, owing to the negligible scattering probability and its exponential decay, the primary source of decoherence at room temperature for most dopant concentrations is undoubtedly the electron density. This finding strongly suggests that electron-electron scattering can influence spin lifetime only through its tendency to randomize the carriers' energy. The energy randomization, in turn, has a significant impact on other scattering mechanisms, in particular, on the  Fr\"{o}hlich  scattering, which yields a substantial contribution to the spin relaxation time. Interestingly enough, this is analogous to the role that e-e scattering plays in the case of electron mobility.~\cite{pierret1987}.

Overall, our analysis reconciles the semiclassical Monte Carlo approach with the computational first-principles method proposed by XHSP. Furthermore, as a by-product of our work, we discovered that at dopant concentrations of about \SI{2.5}{\times 10^{17}\cm^{-3}}, the approximate—i.e., computed in 1BA—and exact electron-electron cross-sections, the latter being obtained by accurately solving the radial  Schr{\"o}dinger equation, differ by a magnitude similar to those calculated in alkali atoms ~\cite{kukkonen1973}.

 \section{Modelling the Interactions and Collisions in $n$-doped Gallium Arsenide} \label{interactions}

\begin{figure}
\centering
\includegraphics[scale=1.]{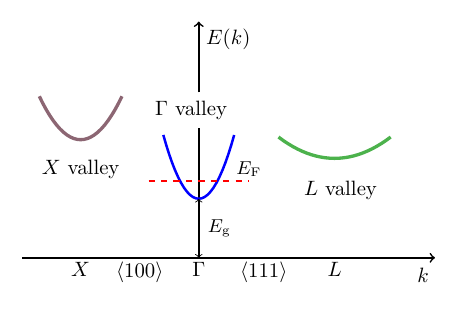}
\caption{Schematic picture of the conduction band valleys in zincblende GaAs. The dashed line corresponds to the Fermi energy $E_{\rm F}$. In the present work the  energy is set equal to zero at the bottom of the $\Gamma$ valley. Note that  the energy gap is $E_{\mathrm{g}} = 1.424$ $\mathrm{eV}$ at $T= 300$ $\mathrm{K}$~\cite{bernardi2015}.}
\label{fig:bandstructureScheme}
\end{figure}

The band structure of  GaAs is constituted by three kind of valleys denoted by the symbols  $\Gamma, L, X$ according to the corresponding crystal symmetry points, see Fig.~\ref{fig:bandstructureScheme}. In EMC it is assumed that 
the carriers are confined at the bottom of a parabolic $\Gamma$ valley~\cite{marchetti2014, marchetti2014a}, and so only intravalley collisions are possible. As a result, there exists a quadratic energy dispersion between the carrier's energy $E$ and its wave vector magnitude $k$, i.e. $E \sim k^{2}$. 

During their  dynamics, the carriers can collide according to  different scattering mechanisms~\footnote{According to the EMC algorithm the carriers can also undergo a fictitious scattering process referred to as the self-scattering,  where nothing actually happens.}. The electrons scatter off  the longitudinal acoustic  phonons  and longitudinal optical phonons, whose corresponding interaction potentials  are modelled according to the deformation potential and Fr\"{o}hlich Hamiltonian~\cite{jacoboni1983, feliciano2017}, respectively. 
Due to the Coulomb interaction there may be two other possible types of collisions:  the e-e scattering and the electron-impurity (e-i) scattering. For the latter it is assumed that  the dopant concentration $n_i = n_e$ as in Ref.~\cite{xu2021}. Regarding the Coulomb interaction, the exceptionally challenging issue lies in accurately modeling the corresponding potential while accounting for dielectric screening effects. To this end, XHSP adopt the GW approximation to model the screened Coulomb interaction~\cite{reining2018, golze2019}, neglecting the exchange part because they found it was important only for the electron-hole exchange mechanism  in heavily doped $p$-type materials. 
In the EMC approach it is customary to tackle the above problem starting from the random phase approximation~\cite{lindhard1954, giuliani2005, pines2016}. The RPA  should  be a suitable approximation for the Coulomb potential in free electron-like materials such as GaAs~\cite{enfietzoglou2013}
where short-range exchange and correlation effects can be neglected. In principle, its  validity can be supported by looking at the smallness of the density parameter (or the Wigner-Seitz radius)  $r_s$  in units of the effective Bohr radius $a_0^{\ast} = 4 \pi \varepsilon \hbar^{2}/m^{\ast} e^{2} $ where $\varepsilon, \hbar, m^{\ast}, e$ are the material dielectric constant, the reduced Planck constant, the effective mass of the electron and the magnitude of the electron charge, respectively. In our case, it is found that the Wigner-Seitz radius $r_s$ ranges from $\approx 0.9$  to $ \approx 2.8$, assuming that  $\varepsilon=12.9\varepsilon_{0}$ and $m^{\ast} = 0.067 m_{\rm{el}}$ where $\varepsilon_{0},  m_{\rm{el}}$ are the permittivity of the vacuum and the electron mass, respectively ~\cite{marchetti2014, vurgaftman2001}.  This finding strongly supports the use of the RPA for the problem at hand~\cite{enfietzoglou2013}.

Furthermore, within the EMC  the effective RPA Coulomb static interaction stems from the self-consistent static dielectric function $\varepsilon^{\rm RPA}\left(\mathbf{q}, \omega =0\right)$ in the long-wavelength limit, i.e. for $|\mathbf{q}|  \approx 0$. This further approximation yields the Thomas-Fermi potential~\cite{hohenberg1964} (or equivalently the  Debye-H{\"u}ckel potential~\cite{debye1923, chattopadhyay1981}). The latter  takes the form $V\left(r\right)= V_0\exp\left(-q_0 r\right)/r$ where $V_0 = \pm \left(e^{2}/4 \pi \varepsilon \right) $  and the corresponding positive (negative) sign refers to the repulsive  e-e (e-i attractive)  interaction, in the real space. 
The Thomas-Fermi screening parameter (also called inverse screening length) $q_0$ at finite temperature is given by the following formula: $q_0= \left(n_e e^{2}/ \varepsilon k_{\mathrm{B}}T\right)  \mathscr{F}_{-1/2}(\eta)\mathscr{F}_{1/2}^{-1}(\eta)$~\cite{chattopadhyay1981} where  $k_B, \eta,  \mathscr{F}_{j}$ denote the Boltzmann constant, the reduced electronic chemical potential~\footnote{The reduced electronic chemical potential is here defined by $\eta=\mu/\left(k_B T\right)$ where $\mu$ is the electron chemical potential.} and the  Fermi-Dirac integral of order $j$~\cite{chattopadhyay1981, Koroleva2017}, respectively. Note that the consistency of the TF potential has been proved valid for the material parameters under scrutiny~\cite{marchetti2014}.

Once the screened interaction potential is given, one can compute the frequency $w_{\rm ee}$ at which the e-e collisions occur through the Fermi's Golden rule ~\cite{Ashcroft76}:
\begin{equation}\label{eq:fgr}
w_{\mathrm{ee}}  = \frac{ 2 \pi}{\hbar} n_e |\Bra{\mathbf{k'}} V \Ket{\mathbf{k}}|^{2} \delta \left(E'-E\right) \, ,
\end{equation}
where $\mathbf{k}$, $E$ and $\mathbf{k}^{\prime}$, $E'$  denote the carrier's wave vectors and energies before and after a collision, respectively.  It is worth noting that the e-e scattering rate given by Eq.~\ref{eq:fgr} is valid only if the carries themselves  constitute  scattering centers sufficiently dilute. Also,  Eq.~\ref{eq:fgr}  neglects the Pauli blockade, and hence $w_{\rm ee}$ is independent of the electronic distribution. Accordingly, it is found from the EMC simulations that  the carriers form a Maxwellian gas (or a nondegenerate electron gas) for the materials parameters under scrutiny~\cite{marchetti2014}.

Finally, the FGR entails the first Born approximation ~\cite{fischetti2016}. As a result, one can show that $w_{\rm ee} \sim n_e  \sigma_{\rm ee}^{\rm B}$ up to a constant, where  $\sigma_{\rm ee}^{\rm B}$ denotes the total cross-section for e-e scattering in 1BA~\cite{povh2015}. 
The latter  yields the probability of scattering at a given carrier's collision energy.



\section{Electron-electron Scattering: Results and Discussion} \label{results}


Due to the lack of spatial inversion symmetry in GaAs a spin-orbit interaction (SOI) arises from the band splitting that removes the spin degeneracy~\cite{fabian2007}. This SOI is then described by an effective Hamiltonian, the Dresselhaus Hamiltonian $H_{\mathrm{D}}=\hbar  \mathbf{\Omega}(\mathbf{k})\cdot \vec{\mathbf{\sigma}}$ where the symbol $\vec{\mathbf{\sigma}}=\left(\sigma_x,\sigma_y,\sigma_z\right)$ denotes a vector whose components are the Pauli matrices, and $\mathbf{\Omega}(\mathbf{k})$ is the Larmor precession frequency vector whose expression is ~\cite{dresselhaus1955, FU2008}
\begin{equation}\label{eq:larmor}
 \mathbf{\Omega}(\mathbf{k})=\frac{\gamma_{so}}{\hbar}[k_{x}(k_{y}^2-k_{z}^2), k_{y}(k_{z}^2-k_{x}^2), k_{z}(k_{x}^2-k_{y}^2)] \, ,
\end{equation}
where $k_{i}$ and  $\gamma_{so}$ are the wave vector components along the cubic crystal axes ($i=x,y,z$) and the spin-orbit coupling, respectively. As a result, the electron spin precesses along a momentum-dependent magnetic field. During each collision event, this magnetic field causes the precession to alter its direction and frequency. We refer the reader to Refs.~\cite{marchetti2014, marchetti2014a} for details about the implementation of the corresponding spin dynamics within the EMC.

\begin{figure}
\centering
\includegraphics[scale=0.55]{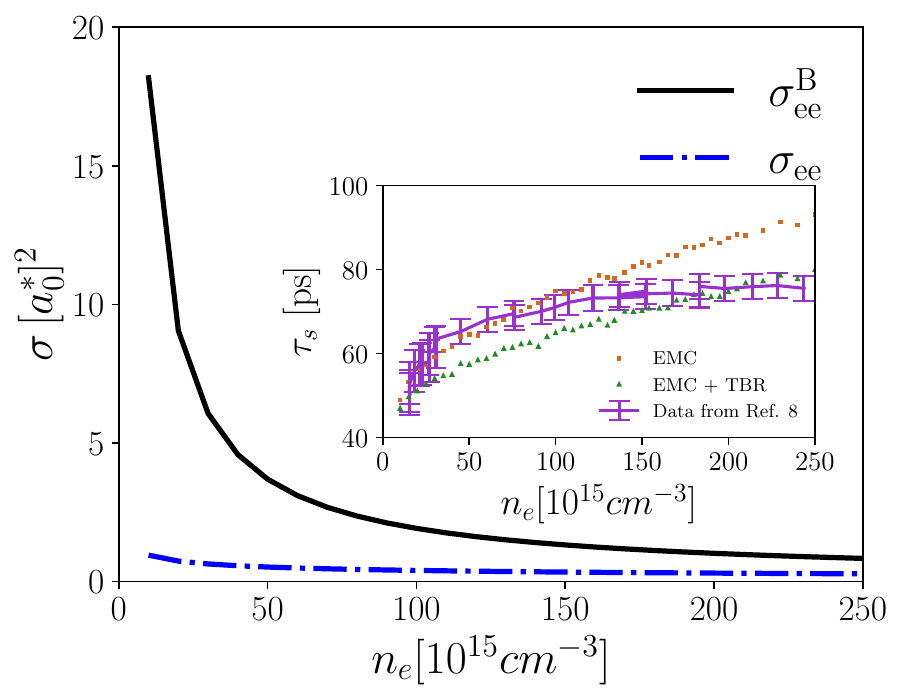}

\caption{The electron-electron total cross section $\sigma_{\rm ee}$ either in the first Born approximation ($\sigma_{\rm ee} ^{\rm B}$)  or exact  $\sigma_{\rm ee}$ ($l_{\rm max} =4$) as function of the electron density $n_e$ computed at the carrier's average thermal energy.  (Inset) Spin relaxation times $\tau_s$ at room temperature as functions of $n_e$ obtained from the EMC simulations  with  the TBR method (triangle symbols) and without it  (square symbols) using  $N= 25,000$ carriers and assuming $\gamma_{\rm so}= \SI{21.9}{\electronvolt \angstrom ^3}$ together with the experimental data from  Ref.~\cite{oertel2008}. The error bars in the numerical simulations are not displayed due to their small size.}
\label{fig:cross_section}     
\end{figure}

Next, we briefly recall the main findings about the spin relaxation time obtained through the EMC simulations after the system reaches the thermal equilibrium~\cite{marchetti2014, marchetti2014a}. Note that the EMC simulations were performed   with $N = 25 \times 10^{3}$ carriers and assuming   the spin-orbit coupling parameter $\gamma_{\rm so}= \SI{21.9}{\electronvolt \angstrom ^3}$~\cite{fabian2007, oertel2008, FU2008}.  In  Fig.~\ref{fig:cross_section} (inset) the values of the spin relaxation time  $ \tau_{s}^{\rm EMC}$ (square symbol) obtained from the numerical simulations  are compared with the  data obtained through the photoluminescence spectroscopy experiments~\cite{oertel2008}  with  dopant concentration $n_i$ ranging from \SI{}{10^{16}\cm^{-3}} to $2.5 \times  10^{17}$ cm$^{-3}$ at $T=\SI{300}{\kelvin}$. It is found that within the experimental errors, $ \tau_{s}^{\rm EMC} \approx  \tau_{s}^{\rm EXP}$ for $n_e \lesssim \SI{1.3}{\times 10^{17}\cm^{-3}} $, however, shortly thereafter the numerical simulations start to overestimate the spin lifetime. This discrepancy becomes more pronounced as the electron density increases, strongly indicating that the FGR fails to provide accurate scattering rates for Coulomb interactions. Indeed, as the scattering centers cease to be dilute, rendering the underlying two-body model of FGR invalid, addressing the complex and unresolved issue of multiple scattering becomes necessary, see Ref.~\cite{joachain2023} and references therein. To this end, Ridley introduced the third-body rejection (TBR) method~\cite{ridley1977, roer1986}, as a pragmatic approach to tackle this issue~\footnote{The TBR aims to quantify the assumption that only one centre is active during the scattering process, by computing the probability that no second centre can interfere.}. In  Fig.~\ref{fig:cross_section} (inset) the spin relaxation time  $ \tau^{\rm EMC + TBR} $ (triangle symbol) from the numerical simulations equipped with TBR applied to the e-e scattering, is shown as function of $n_e$. It is found that $ \tau^{\rm EMC + TBR} \approx  \tau^{\rm EXP}$ for $n_e \gtrsim \SI{1.5}{\times 10^{17} \times \cm^{-3}} $, thus confirming the excellent agreement between the  numerical and experimental results in the region where the standard EMC approach proves to overestimate the experimental spin lifetimes. 

Nevertheless, the above results do not tell us what is the actual role of the electron-electron interactions in the spin decoherence. In this regard, first we note that the curves relative to  $ \tau^{\rm EMC} $ and  $ \tau^{\rm EMC + TBR} $  increase monotonically  with $n_e$. This electron density-dependence is certainly dictated by the Coulomb interactions because in the EMC approach the electron-phonon scattering rates do not depend upon $n_e$ directly. Furthermore, we can state that such a  dependence is linear. To illustrate this fact, we use the linear regression as the simplest best fit of  $ \tau^{\rm EMC} $ and $ \tau^{\rm EMC + TBR}$ according  to the Bayesian Occam's razor~\cite{jefferys1992, Ghahramani2015}. This means that we are aware that  the curve fitting can be further improved by increasing the number of the parameters, e.g. by fitting the four parameter logistic (4PL) curve to the data.  However,  as the above models, i.e. the linear and the nonlinear 4PL, fit equally well with  the respective coefficients of determination $R^{2}$ close to unity, the simpler model is very likely to be a better description of the underlying physics for the problem at hand. This is indeed an approach  similar to  that  used by Galileo for deriving  the law of the falling bodies~\cite{jefferys1992}.
 
\begin{figure}
    \centering
    \begin{minipage}{0.5\textwidth}
        \includegraphics[width=\linewidth]{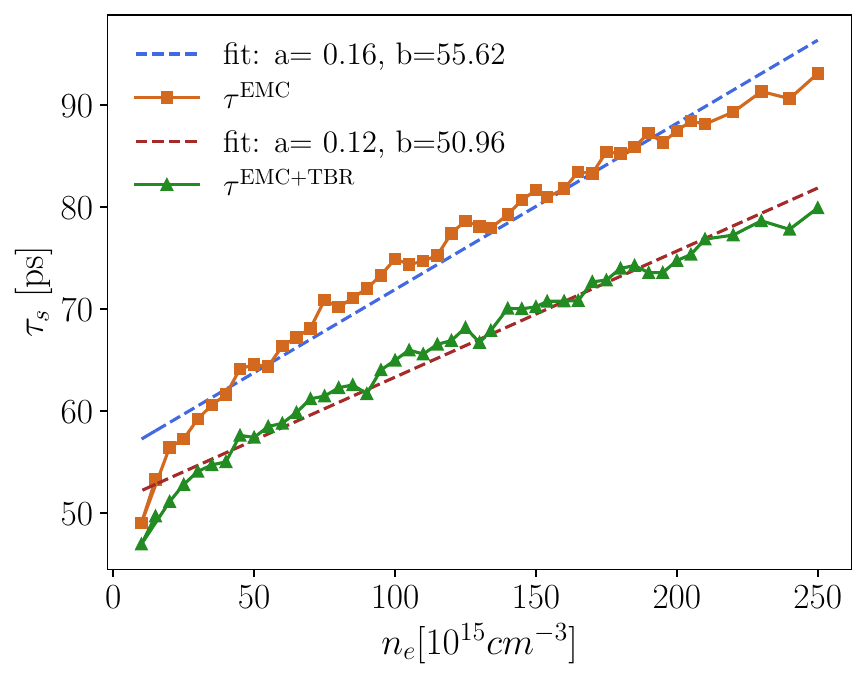} 
        \caption{Fit of the linear  model 
        $\tau_s =  a n_e + b$ to data relative to  $ \tau^{\rm EMC} $ and $ \tau^{\rm EMC + TBR}$ . The corresponding    coefficients of determination are $R^{2} \approx 0.96 $ and $R^{2} \approx 0.97 $, respectively.}
        \label{fig:fitTau}
    \end{minipage}
    \hfill
    \begin{minipage}{0.5\textwidth}
        \includegraphics[width=\linewidth]{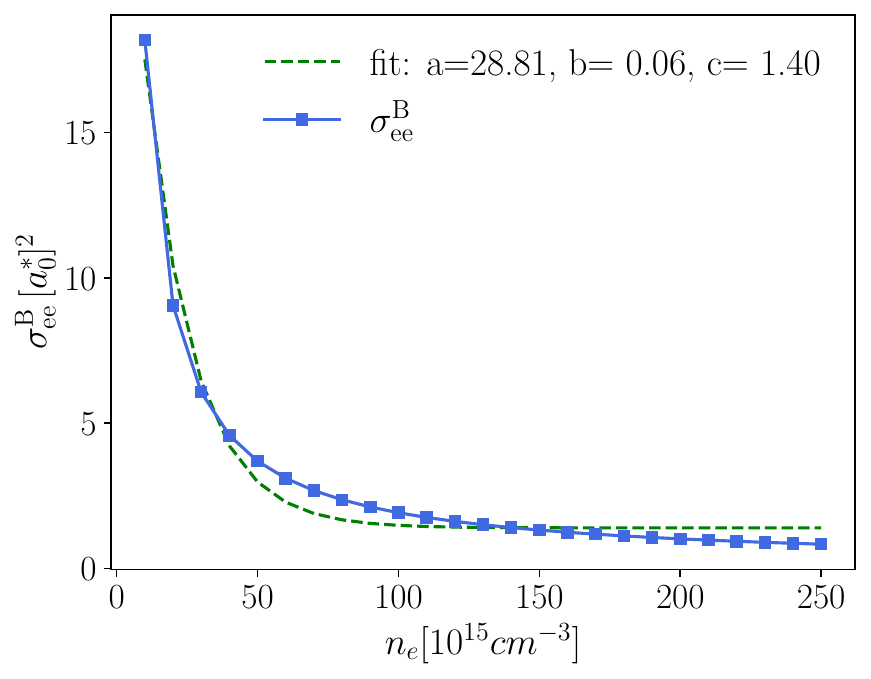}
        \caption{ Fit of the exponential  model  $ \sigma_{\rm ee} ^{\rm B} =  a \exp(-b n_e) + c$ 
        to the scattering data relative to $\sigma_{\rm ee} ^{\rm B}$. The corresponding    coefficient of determination is $R^{2} \approx 0.98 $.}
        \label{fig:fitSigma}
    \end{minipage}
\end{figure}

 Our results of the curve fitting using the liner model $\tau_s =  a n_e + b$  together with the respective estimate values of $a, b$ are shown in Fig.~\ref{fig:fitTau}  (see also Table~\ref{table:table1}). The  goodness of such fits is confirmed by the fact that   $R^{2} \approx 0.96 $ and $R^{2} \approx 0.97 $  for the EMC and EMC + TBR data, respectively. However, a relatively small departure from the linear model is observed at low dopant concentrations in both cases. In the following, we shall argue that this is the only possible direct effect of the electron-electron scattering on the spin relaxation time. Thus, disregarding the e-i interactions~\footnote{The e-i collisions lead to the very same electron density-dependence for  $ \tau^{\rm EMC} $~\cite{marchetti2014}.}, we can conclude that the spin relaxation time  due to the e-e interactions grows linearly with $n_e$, that is, $ \tau_{\rm ee}^{\rm EMC} \sim n_e$, $ \tau_{\rm ee}^{\rm EMC + TBR} \sim n_e$, where $ \tau_{\rm ee}^{\rm EMC} \approx  \tau^{\rm EMC} \left(n_e\right)$,  $ \tau_{\rm ee}^{\rm EMC + TBR} \approx  \tau^{\rm EMC + TBR} \left(n_e\right)$,

The remaining crucial question is what is the dependence upon $n_e$ of the e-e scattering probability  $ \sigma_{\rm ee} ^{\rm B}$. To begin with, we note that the nonrelativistic total electron-electron cross-section  $ \sigma_{\rm ee} ^{\rm B}$ must be calculated in the center of mass (CoM). Hence, such a scattering problem is equivalent to that of a fictitious particle with kinetic energy $E=1/2 \mu v_{\rm r}^{2}$, where $\mu = 0. 5 m^{\ast}$, $v_r$ are the reduced mass and the relative speed, respectively, scattering off the Thomas-Fermi potential.  As the spin lifetime is obtained at thermal equilibrium~\cite{marchetti2014}, in good approximation one can assume that the typical collision velocity corresponds to the thermal average of relative velocity, here denoted by  $\Bar{v}_{\rm r}$. Then,  by means of the Maxwell-Boltzmann distribution one can show that  $\Bar{v}_{\rm r} = \sqrt{2} \Bar{v} $ where  $\Bar{v} = \left(8k_{B}T/\pi m^{\ast}\right)^{\frac{1}{2}}$ is the average speed~\cite{henriksen2008}. Then, by means of  the quadratic energy dispersion, see Section~\ref{interactions}, one finds that in the CoM  $k =  \mu \Bar{v}_{\rm rel}/\hbar $ yields the
magnitude of the wave vector of the incoming plane wave to be used as input in the following formula for the total electron cross section in 1BA~\cite{galitski2013} 
\begin{equation}\label{eq:total_0}
 \sigma_{\rm ee} ^{\rm B} \left(k\right) = 16 \pi \left(\frac{\mu V_0}{\hbar^{2}}\right)^{2}  \frac{1}{q_{0}^{2}\left( q_{0}^{2} + 4k^{2}\right)}\, .
\end{equation}

In Fig.~\ref{fig:cross_section} we plot $\sigma_{\rm ee}$ (black solid line) 
computed according to Eq.~\ref{eq:total_0} in units of $[a_0^{\ast}]^{2}$, as a function of $n_e$ for the material parameters under scrutiny. We note that the corresponding curve exhibits a monotonically decreasing trend as  $n_e$ increases. This is clearly in contrast with the observed behaviours of  $ \tau^{\rm EMC}$ and $  \tau^{\rm EMC + TBR}$. Again, according to to the Occam's razor, we proceed to fit the following exponential curve $ \sigma_{\rm ee} ^{\rm B} =  a \exp(-b n_e) + c$   to the above  scattering data.  Such a fit is shown in Fig.~\ref{fig:fitSigma}, see also Table~\ref{table:table1}). Again, the goodness of such a curve fitting is confirmed by the fact that $R^{2} \approx 0.98$. 

Thus, by comparison of the monotonic trends of $ \tau_{\rm ee}^{\rm EMC} \sim n_e, \tau_{\rm ee}^{\rm EMC + TBR} \sim n_e$ with the exponential decay of $  \sigma_{\rm ee} ^{\rm B} \sim  \exp(-0.06 n_e) $ one can certainly conclude that  the electron density itself is the main source of spin relaxation time, as already  pointed out in Ref.~\cite{D’Amico_2019}, except for the low dopant concentrations close to   \SI{}{10^{16}\cm^{-3}}  where the magnitude of scattering probability $  \sigma_{\rm ee} ^{\rm B} $ increases substantially, possibly leading to a departure from linearity  of the fits for  $ \tau_{\rm ee}^{\rm EMC}, \tau_{\rm ee}^{\rm EMC + TBR}$.

\begin{table*}
\renewcommand{\arraystretch}{2} 
\caption{The  optimal values of parameters $a, b, c$ obtained  by fitting the curves  $\tau_s =  a n_e + b$ and   $ \sigma  =  a \exp(-b n_e) + c$     to the data relative to spin relaxation times  $ \tau^{\rm EMC} , \tau^{\rm EMC + TBR}$  and the electron-electron total cross sections   $\sigma_{\rm ee} ^{\rm B}, \sigma_{\rm ee}$  (the cut-off value is $l_{\rm max} =4$ for the latter), respectively. In the last column the corresponding values of the coefficient of determination $R^{2}$. }
\label{table:table1}
  \begin{ruledtabular}
    \begin{tabular}{c | c   c   c  c c c  }
   Data &  $a$  & $b$ &   $c$ &   $R^{2}$ \\ [1ex]
    \hline
    $ \tau^{\rm EMC}$    &  $0.16 \pm  0.004$    & $55.62 \pm 0.68 $        & $-$      & $0.96$          
    \\
    
 $  \tau^{\rm EMC + TBR}$    &   $0.12 \pm 0.003$    & $50.95 \pm 0.46$        & $-$      & $0.97 $        \\
 
  $\sigma_{\rm ee} ^{\rm B}$      &   $28.81 \pm 1.74 $   & $0.06 \pm 0.004$        &  $1.40 \pm 0.13$       &  $0.98  $     \\

    $\sigma_{\rm ee}$    &  $0.75 \pm 0.03 $     &       $0.02 \pm 0.001 $   &    $0.31 \pm 0.01 $      &    $0.98 $    \\

    \end{tabular}
\end{ruledtabular}
\end{table*}
%

In the following, we shall attempt to understand the physical origin of this exponential decay in the scattering probability. To achieve this, 
we shall computed the electron-electron exact cross section $\sigma_{\rm ee}$ as function of the electron density by accurately solving   the radial  Schr{\"o}dinger equation. There are  two  reasons for doing so. First, we wish to get  a further confirmation of the exponential decay by means of an alternative approach that in this case goes beyond the Born approximation. Note that in general going beyond the 1BA is not always an improvement; in fact, it depends on the specific system, especially in condensed matter systems~\cite{shankar}. Nevertheless, it always useful to compare results obtained from different scattering methods. Second, the behavior of the phase shifts as a function of the interparticle distance sheds light on the physical origin of the observed exponential trend, which is not easily seen from  Eq.~\ref{eq:total_0}) in 1BA.

In the CoM frame, the  exact  e-e  cross section $\sigma_{\rm ee}$ is obtained  by computing the phase shifts $\delta_l$, where   $l = 0, 1, \dots$ denotes the angular momentum number. For spherically symmetric potentials $\sigma_{\rm ee}$ reads~\cite{bethe1957}
\begin{equation}\label{eq:total}
\sigma_{\rm ee}\left(k\right) = \frac{4 \pi}{k^{2}} \sum_{l=0}^{\infty} \left(2l +1 \right) \sin^{2} \delta_l \, .
\end{equation}

In general, the series in Eq.~\ref{eq:total} converges after a finite number of partial waves. Here, we shall denote the maximum number of partial waves required for the numerical convergence as $l_{\rm max}$. The computations of $\delta_l$ require solving the Schr{\"o}dinger  equation. To this end, we shall employ the variable phase method (VPM)~\cite{calogero1963, babikov1967, marchetti2019}. The VPM utilizes the following first-order nonlinear equation, akin to a generalized Riccati equation:

\begin{multline}\label{eq:phase}
  \delta'_l \left(r \right) = - \frac{ 2 \mu V\left(r \right) }{ \hbar^{2} k} \\ \times \left[\cos\delta_l\left(r \right)	\hat{j_l} \left(kr \right)  - \sin\delta_l\left(r \right)\hat{n}_l \left(kr \right)  \right]^{2}\, , 
\end{multline}
where 	$\hat{j_l}$, $\hat{n}_l$ are the Riccati-Bessel functions. The boundary condition at the origin is  $\delta_l\left(0\right)=0$~\footnote{The numerical values of $\delta_l$ are then obtained asymptotically, i.e. taking the following limit $\lim\limits_{r\to \infty} \delta_l\left(r\right)$. }. Note that we found that five  partial waves are sufficient, i.e.  $l_{\rm max} =4$, to ensure the numerical convergence of the series  in  Eq.~\ref{eq:total} for the problem at hand.  

In Fig.~\ref{fig:cross_section}, $\sigma_{\rm ee}$ (blue dash-dot line) as function of  $n_e$ is shown. It exhibits the very same electron density dependence of $\sigma_{\rm ee} ^{\rm B}$, 
although this similarity might not be immediately noticeable due to the substantial differences in magnitudes between the cross sections.  In fact, at low dopant concentrations   $\sigma_{\rm ee}^{\rm B} \approx 19 \sigma_{\rm ee}$.  Nevertheless, when fitting the exponential curve to this  scattering data set, we obtain an excellent fit (not shown) with the coefficient of determination   $R^{2} \approx 0.98$ (we refer the reader to  Table~\ref{table:table1} for the optimal values of the fit). 
Interestingly enough,  $\sigma_{\rm ee}^{\rm B} \approx 3 \sigma_{\rm ee}$ at   \SI{2.5}{\times 10^{17}\cm^{-3}}, a result similar to what found by Kukkonen and Smith  for the electron-electron collisions in  alkali metals where  $\sigma_{\rm ee}^{\rm B} \approx 2 \sigma_{\rm ee}$~\cite{kukkonen1973}.

\begin{figure} 
\resizebox{0.50\textwidth}{!}{%
  \includegraphics{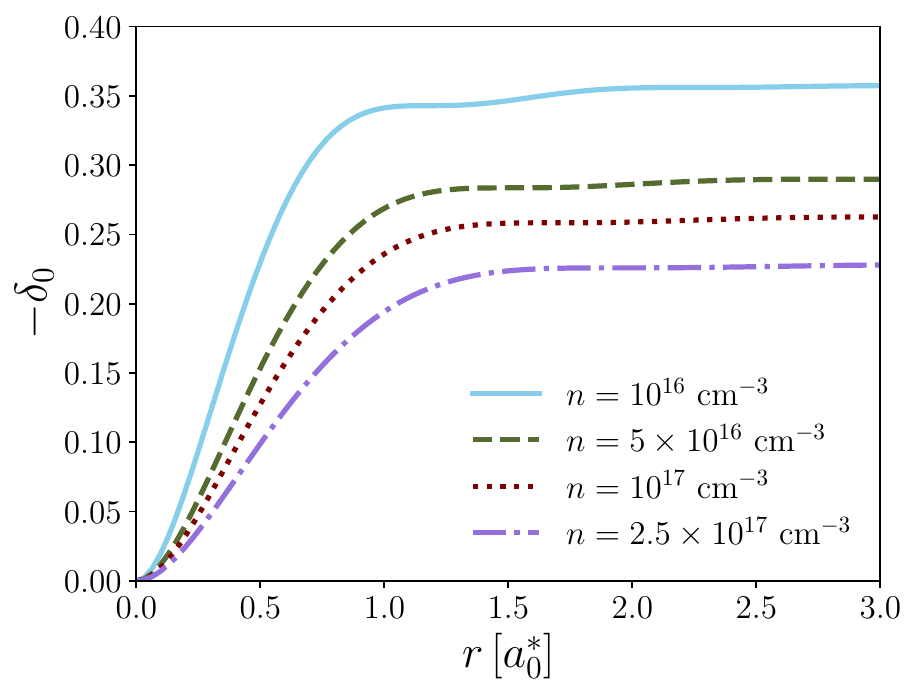}
}
\caption{The  s-wave phase shift $-\delta_0$ as function of the interparticle distance $r$ in units of the effective Bohr radius $a_0^{\ast}$ due to the repulsive electron-electron interaction for the dopant concentrations:  \SI{}{10^{16}\cm^{-3}},  \SI{5}{\times 10^{16}\cm^{-3}}, \SI{}{10^{17}\cm^{-3}}  and \SI{2.5}{\times 10^{17}\cm^{-3}} . The corresponding asymptotic phase shift $\delta_0$ takes the values $\approx -0.35$,  $\approx -0.29$,  $\approx -0.26$ and  $ \approx -0.22$, respectively.}
\label{fig:phase_shift}       
\end{figure}

The monotonically decreasing trend and the convergence of the exact and approximate cross sections at high dopant concentrations can now be clearly understood by means of the scattering phase shifts. In this regard, we will focus on studying the phase shift in s-wave ($l=0$) $\delta_0$, which contributes significantly to the scattering probability. In Fig.~\ref{fig:phase_shift},  $-\delta_0$~\footnote{Note that the phase shift $\delta_0$ is negative because the electron-electron interaction potential is repulsive.}  as function of the interparticle distance $r$ in units of $a_0^{\ast}$ is shown   for four different dopant concentrations: \SI{}{10^{16}\cm^{-3}}, \SI{5}{\times 10^{16}\cm^{-3}}, \SI{}{10^{17}\cm^{-3}}  and \SI{2.5}{\times 10^{17}\cm^{-3}}. The corresponding curves of the phase function, i.e.  $r \rightarrow \delta_0\left(r\right)$, at low dopant concentration yield larger absolute asymptotic values with respect to those observed at high dopant concentrations. The reason for this is that at low dopant concentrations, the range of the Thomas-Fermi  potential increases significantly, thereby affecting the electron's motion over longer distances away from the carrier target.

Finally, we note that the monotonically decreasing trend of  $-\delta_0$ implies that the Born approximation improves as the electron density increases. In fact, when the exact phase shifts become much smaller of unity (or equivalently of $\pi/2$ according to Ref.~\cite{bethe1957}), one expects that  $\delta_{l}^{B} \approx \delta_{l}$~\cite{bethe1957} for each value of $l$~\cite{rau1976}. As a consequence, there must be a convergence of the approximate and exact scattering cross sections, i.e.,  $\sigma_{\rm ee}^{\rm B} \approx  \sigma_{\rm ee}$ increasing $n_e$. However, we should not ignore that   the Born approximation can do very well, even  when the phase shifts are not very small~\cite{rau1976}

\section{Conclusion} \label{conclusion}

There is no doubt that electron-electron  interactions play an important role in condensed matter systems. For instance, they determine the quasiparticle lifetime in a Fermi liquid~\cite{giuliani2005}. However, they do not directly affect electron mobility because scattering between electron pairs conserves the total momentum of the system~\cite{pierret1987}. Nevertheless, electron-electron scattering randomizes the distribution of the electronic total momentum. This leads to an indirect second-order effect on mobility, influencing it only through any energy dependence of other scattering mechanisms, such as electron-phonon collisions~\cite{MCLEAN1960220}.

The analysis presented here is consistent with the above physical picture, albeit in the context of spin lifetime due to the Dyakonov-Perel mechanism. The fact that we have demonstrated the main source of spin decoherence is electron density implicitly reaffirms the crucial role of electron-electron scattering in providing the second-order effect necessary to influence spin lifetime, a point that was previously suggested in Ref.~\cite{marchetti2014}.

\begin{acknowledgments}
I would like to thank Yuang Ping and Junqing Xu for their detailed explanations of the physics and electron-electron interaction in accordance with their \emph{ab initio} approach, and Fabio Caruso for his insightful comments on the GW approximation. Furthermore, I take this opportunity to thank Irene D'Amico and Matthew Hodgson for their past collaboration in the field of spintronics.
\end{acknowledgments}


\bibliography{references}{}

\bibliographystyle{apsrev4-1}

\end{document}